\documentclass{ifacconf}

\usepackage{amsmath,amssymb,graphicx,natbib}
\usepackage{xcolor}
\usepackage{tikz}
\usetikzlibrary{arrows.meta,positioning,fit,backgrounds,shapes.geometric}
\usepackage{listings}
\usepackage{capt-of}
\usepackage{booktabs}

\definecolor{mimosmeas}{RGB}{140,140,140}
\definecolor{mimoss1}{RGB}{0,114,189}
\definecolor{mimoss1res}{RGB}{86,180,233}
\definecolor{mimoss2}{RGB}{230,115,25}
\definecolor{mimoss2res}{RGB}{240,180,120}
\definecolor{mimosnoise}{RGB}{51,166,77}
\definecolor{matlabgreen}{RGB}{0,128,0}
\definecolor{matlabblue}{RGB}{0,0,255}
\definecolor{matlabpurple}{RGB}{163,21,21}

\lstdefinestyle{matlab}{
    language=Matlab,
    basicstyle=\ttfamily\scriptsize,
    keywordstyle=\color{matlabblue},
    commentstyle=\color{matlabgreen},
    stringstyle=\color{matlabpurple},
    morestring=[b]",
    numbers=none,
    frame=single,
    rulecolor=\color{black!45},
    framerule=0.4pt,
    framesep=3pt,
    backgroundcolor=\color{black!1},
    breaklines=true,
    columns=fullflexible,
    keepspaces=true,
    showstringspaces=false,
    aboveskip=0.6em,
    belowskip=0.4em
}

\definecolor{fbact}{RGB}{0,90,180}
\definecolor{fbsens}{RGB}{235,95,15}

\definecolor{oatactz}{RGB}{0,90,180}
\definecolor{oatactx}{RGB}{0,165,205}
\definecolor{oatacty}{RGB}{105,75,190}

\definecolor{oatsensz}{RGB}{235,95,15}
\definecolor{oatsensx}{RGB}{195,45,35}
\definecolor{oatsensy}{RGB}{235,155,15}

\definecolor{oatgrav}{RGB}{65,145,95}

\DeclareRobustCommand{\tikzline}[1]{%
  (\tikz[baseline=-0.6ex,x=1pt,y=1pt,line width=0.7pt]{%
    \draw[draw=#1] (0,0) -- (10,0);})%
}

\makeatletter
\let\hyper@nopatch@sectioning\@empty
\makeatother
\usepackage[hidelinks]{hyperref}

\usepackage{dblfloatfix}

\usepackage{eso-pic}
\AddToShipoutPictureBG*{%
\AtPageUpperLeft{%
\setlength\unitlength{1in}%
\hspace*{\dimexpr0.5\paperwidth\relax}
\makebox(0,-1.75)[c]{
\begin{tabular}{c c}
Maarten van der Hulst \emph{et al.},
MIMOS: A Toolbox for Multivariable \\
Modal Identification, uploaded to ArXiv  \today \\
\end{tabular}}}}

\AddToShipoutPictureBG*{%
\AtPageUpperLeft{%
\setlength\unitlength{1in}%
\hspace*{\dimexpr0.5\paperwidth\relax}
\makebox(0,-23.2)[c]{
\footnotesize
\begin{tabular}{c c}
This work has been submitted to IFAC for possible publication.
\end{tabular}}}}

\begin{document}
\begin{frontmatter}

\title{MIMOS: A Toolbox for Multivariable Modal Identification\thanksref{funding}}

\thanks[funding]{This project is funded by Holland High Tech | TKI HSTM via the PPP Innovation Scheme (PPP-I) for public-private partnerships.}

\author[TUE]{M. van der Hulst}
\author[TUE]{R. A. Gonz\'alez}
\author[TUE]{K. Classens}
\author[TUE]{P. Tacx}
\author[ASML]{N. Dirkx}
\author[ASML]{J. van de Wijdeven}
\author[TUE]{T. Oomen}

\address[TUE]{Department of Mechanical Engineering, Eindhoven University of Technology, The Netherlands}
\address[ASML]{ASML, Veldhoven, The Netherlands}

\begin{abstract}
Accurate and physically interpretable models are essential for the analysis and control of complex multivariable systems.
This paper presents the Multivariable Identification of Modal Optical and mechatronic Systems (MIMOS) toolbox, an open-source MATLAB toolbox for identifying physical modal models from measured multivariable frequency response functions.
The two-stage approach first estimates an additive input-output model from measured frequency responses and subsequently recovers the physical modal structure through information-weighted projection.
MIMOS implements this approach in a practical and scalable workflow combining frequency response preprocessing, interactive modal initialization and refinement, and support for proportional and general viscous damping.
Experimental results on a publicly available $6\times8$ vibration isolation benchmark and a $7\times17$ prototype wafer stage demonstrate the capability of MIMOS to effectively handle identification problems involving thousands of parameters and estimate accurate, minimal-order physical modal models.
\end{abstract}

\begin{keyword}
System identification; modal analysis; frequency response; software tools; multivariable systems.
\end{keyword}

\end{frontmatter}

\section{Introduction}\label{sec:introduction}
Physically interpretable parametric models are essential for
model-based analysis and control of high-performance optical and mechatronic
systems, with applications ranging from advanced control to mechanical
system design, model validation, and machine monitoring
\Citep{Steinbuch2022MotionLaw}. Modal models are particularly attractive since they provide compact
representations of the dynamics in terms of physically
meaningful modal properties.

Reliably identifying modal models for industrial mechatronic systems is
challenging, since these systems typically have many inputs and outputs
and exhibit complex, lightly damped flexible dynamics
\citep{Steinbuch2022MotionLaw}. Moreover, the physical modal structure imposes low-rank factorizations
on the residue matrices describing the input--output contribution of
each mode \citep{Ewins2000ModalTesting}, resulting in a nonlinear
estimation problem with increasingly many coupled parameters.
To address this complexity, established frequency-domain methods
commonly identify polynomial or pole--residue representations,
possibly followed by nonlinear refinement, from which physical
modal parameters can subsequently be extracted
\citep{Peeters2004TheEstimation,Verboven2005MultivariableEstimation,
Amador2019PoleResidueMaximumLikelihood,
Voorhoeve2021IdentifyingStage,Shirvani2023MIMOIdentification}.
Direct nonlinear methods instead incorporate the physical modal
structure into the frequency-domain identification criterion,
jointly refining modal dynamics and mode-shape parameters
\citep{ElKafafy2013FastMaximumLikelihood,
ElKafafy2016ConstrainedMaximumLikelihood}.
While both approaches enable effective modal identification,
subsequent low-rank reduction can compromise the original
frequency-domain fit
\citep{Peeters2004TheEstimation,Drmac2015VectorFitting},
whereas direct physical estimation requires solving a coupled
nonlinear optimization problem that becomes increasingly
demanding as the number of input--output channels and flexible modes grows.

A two-stage frequency-domain approach for identifying physically structured modal models from Multi-Input Multi-Output (MIMO) frequency-response function (FRF) data was recently introduced in \Citet{HulstModal2026}, with a related time-domain formulation
and statistical analysis presented in
\Citet{Gonzalez2025StatisticallyIdentification}.
The approach first estimates the input--output dynamics without
enforcing the physical modal structure, which is subsequently
recovered through a projection weighted by the information
matrix of the initial estimate. This provides a computationally tractable and statistically informed
route from measured MIMO FRF to physically structured modal models, particularly suited to
systems with many input--output channels and closely spaced flexible modes.

Although the two-stage identification approach has been established
theoretically, its practical application remains limited by the
lack of an integrated, publicly available toolbox implementation.
The aim of this paper is to present MIMOS, an open-source MATLAB
toolbox for identifying physically structured modal models from
MIMO FRF data.
The main contribution is a complete identification workflow
integrating FRF preprocessing and weighting, interactive modal
initialization and refinement, and two-stage estimation,
together with its demonstration on an active vibration isolation
benchmark with open source data and a prototype wafer stage. The MIMOS toolbox, including source code,
documentation, and example datasets, is publicly
available at \url{https://gitlab.tue.nl/m.v.d.hulst/mimos}.

The paper is organized as follows.
Section~\ref{sec:problem} formulates the modal identification problem,
and Section~\ref{sec:two-stage} summarizes the underlying two-stage approach.
Section~\ref{sec:toolbox} presents the MIMOS workflow,
followed by experimental validation in Section~\ref{sec:experiments}
and conclusions in Section~\ref{sec:conclusion}.

\section{Problem formulation}\label{sec:problem}
This section introduces the class of modal models considered in MIMOS
and formulates the corresponding frequency-domain identification problem.

\subsection{Modal models of linear mechanical systems}\label{sec:modal-model}
Consider a linear time-invariant mechanical system described by the
second-order equations of motion
\begin{equation}\label{eq:mechanical-system}
\begin{aligned}
M\ddot q(t)+D\dot q(t)+Kq(t) &= Fu(t),\\
y(t) &= Qq(t),
\end{aligned}
\end{equation}
where $q\in\mathbb R^{n_q}$, $u\in\mathbb R^{n_u}$, and
$y\in\mathbb R^{n_y}$ denote the generalized displacement, input, and
output, respectively. The matrices $M$, $D$, and $K$ denote the mass,
damping, and stiffness matrices, while $F$ and $Q$ describe the
distribution of the inputs and outputs. 
An input--output model of \eqref{eq:mechanical-system} in modal form, retaining
$n_{\mathrm{rb}}$ rigid-body modes and
$n_{\mathrm{flex}}$ flexible modes, is given by
\begin{equation}\label{eq:modal-transfer}
P(s)
=
\frac{\Phi_{\mathrm{l,rb}}\Phi_{\mathrm{r,rb}}^{\mathsf T}}{s^2}
+\sum_{i=1}^{n_{\mathrm{flex}}}P_i(s)
+D_{\mathrm{res}},
\end{equation}
where
$\Phi_{\mathrm{l,rb}}\in\mathbb R^{n_y\times n_{\mathrm{rb}}}$ and
$\Phi_{\mathrm{r,rb}}\in\mathbb R^{n_u\times n_{\mathrm{rb}}}$
contain the rigid-body output and input mode-shape vectors,
respectively, and
$D_{\mathrm{res}}\in\mathbb R^{n_y\times n_u}$ captures the static
contribution of the truncated higher-order flexible modes
\citep{HulstModal2026}. The parameterization of the
flexible modal contributions $P_i(s)$ depends on the damping model.

For proportional damping, the undamped mode shapes also
diagonalize the damping matrix \citep{Ewins2000ModalTesting},
yielding the modal contribution
\begin{equation}\label{eq:proportional-mode}
P_i(s)
=
\frac{\Phi_{\mathrm{l},i}\Phi_{\mathrm{r},i}^{\mathsf T}}
{s^2+2\zeta_i\omega_i s+\omega_i^2},
\end{equation}
where $\omega_i>0$ and $\zeta_i>0$ denote the natural frequency and
damping ratio, respectively. The matrices
$\Phi_{\mathrm{l},i}\in\mathbb R^{n_y\times r_i}$ and
$\Phi_{\mathrm{r},i}\in\mathbb R^{n_u\times r_i}$
contain the real output and input mode shapes, respectively,
with $r_i\geq 1$ modal components. This provides a parsimonious representation, but the
proportional-damping assumption can be restrictive for systems
with more general damping characteristics. To accommodate these systems, MIMOS supports general viscous
damping, for which each flexible mode with $\zeta_i\neq 1$
admits a pole-residue representation of the form

\begin{equation}\label{eq:general-mode}
P_i(s)
=\sum_{k=1}^{2}
\frac{\Psi_{\mathrm{l},i,k}\Psi_{\mathrm{r},i,k}^{\mathsf T}}
{s-\lambda_{i,k}}
=\frac{N_{i,1}s+N_{i,0}}
{s^2+2\zeta_i\omega_i s+\omega_i^2},
\end{equation}
with modal eigenvalues
\begin{equation}\label{eq:modal-poles}
\lambda_{i,1/2}
=
-\zeta_i\omega_i
\pm\omega_i\sqrt{\zeta_i^2-1},
\end{equation}
where $\Psi_{\mathrm{l},i,k}\in\mathbb C^{n_y\times r_i}$
and $\Psi_{\mathrm{r},i,k}\in\mathbb C^{n_u\times r_i}$
contain the output and input mode-shape vectors, respectively.
For underdamped modes with $0<\zeta_i<1$, the eigenvalues
and mode-shape vectors appear in complex-conjugate pairs,
whereas for overdamped modes with $\zeta_i>1$, they appear
in real pairs. For critically damped modes with $\zeta_i=1$,
the eigenvalues coincide, yielding a repeated pole.
The second-order expression in \eqref{eq:general-mode}
remains valid for all $\zeta_i>0$, with real numerator matrices
$N_{i,1},N_{i,0}\in\mathbb R^{n_y\times n_u}$
of rank at most $2r_i$. This parameterization describes a richer class of mechanical
systems at the expense of additional modal parameters.

For both damping models, \eqref{eq:modal-transfer} admits
the real modal state-space realization
\begin{equation}\label{eq:modal-state-space}
\begin{aligned}
\dot x_{\mathrm{m}}
&=
\begin{bmatrix}
0&I\\
-\Omega^2&-2Z\Omega
\end{bmatrix}x_{\mathrm{m}}
+
\begin{bmatrix}
\widetilde{\Phi}_{\mathrm{r}}^{\mathsf T}\\
\Phi_{\mathrm{r}}^{\mathsf T}
\end{bmatrix}u,\\
y
&=
\begin{bmatrix}
\Phi_{\mathrm{l}}&\widetilde{\Phi}_{\mathrm{l}}
\end{bmatrix}x_{\mathrm{m}}
+D_{\mathrm{res}}u,
\end{aligned}
\end{equation}
where $x_{\mathrm{m}}\in\mathbb R^{2n_{\mathrm{m}}}$ and
$n_{\mathrm{m}}=n_{\mathrm{rb}}+\sum_{i=1}^{n_{\mathrm{flex}}}r_i$.
The diagonal matrices
$\Omega,Z\in\mathbb R^{n_{\mathrm{m}}\times n_{\mathrm{m}}}$
contain the modal frequencies and damping ratios, respectively,
repeated $r_i$ times for each flexible mode and zero for
rigid-body modes.
The matrices $\Phi_{\mathrm{l}},\widetilde{\Phi}_{\mathrm{l}}
\in\mathbb R^{n_y\times n_{\mathrm{m}}}$ and
$\Phi_{\mathrm{r}},\widetilde{\Phi}_{\mathrm{r}}
\in\mathbb R^{n_u\times n_{\mathrm{m}}}$
contain the real output and input modal factors, respectively. The numerator matrices in
\eqref{eq:general-mode} are related to the real modal factors
of \eqref{eq:modal-state-space} for general damping through
\begin{equation}\label{eq:general-numerators}
\begin{aligned}
N_{i,1}
&=\Phi_{\mathrm{l},i}\widetilde{\Phi}_{\mathrm{r},i}^{\mathsf T}
+\widetilde{\Phi}_{\mathrm{l},i}\Phi_{\mathrm{r},i}^{\mathsf T},\\
N_{i,0}
&=\Phi_{\mathrm{l},i}\Phi_{\mathrm{r},i}^{\mathsf T}
+(2\zeta_i\omega_i\Phi_{\mathrm{l},i}
-\omega_i^2\widetilde{\Phi}_{\mathrm{l},i})
\widetilde{\Phi}_{\mathrm{r},i}^{\mathsf T},
\end{aligned}
\end{equation}
while for proportional damping
$\widetilde{\Phi}_{\mathrm{l}}=\widetilde{\Phi}_{\mathrm{r}}=0$,
recovering \eqref{eq:proportional-mode}, and with 
$\widetilde{\Phi}_{\mathrm{l,rb}}=
\widetilde{\Phi}_{\mathrm{r,rb}}=0$ for both damping models \Citep{HulstModal2026}.
MIMOS uses \eqref{eq:modal-transfer} as the basis for identification
and returns the corresponding minimal-order, real-valued
state-space model \eqref{eq:modal-state-space}, constructed
directly from the identified physical modal parameters.

\usetikzlibrary{shapes.geometric}

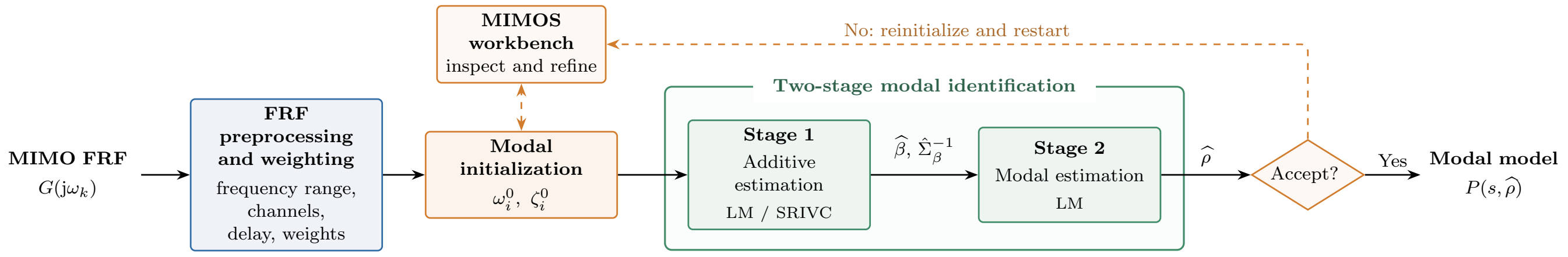
\begin{figure*}[t]
\centering

\resizebox{0.98\textwidth}{!}{%
\begin{tikzpicture}[
    x=1cm,
    y=1cm,
    font=\small,
    >={Stealth[length=2mm]},
    box/.style={
        draw,
        rounded corners=2pt,
        line width=0.7pt,
        align=center,
        minimum height=1.25cm
    }
]


\definecolor{mimosblue}{RGB}{50,110,165}
\definecolor{mimosorange}{RGB}{210,125,45}
\definecolor{mimosgreen}{RGB}{65,140,105}


\node[
    align=center,
    text width=2cm
] (input) at (-0.35,0) {
    \textbf{MIMO FRF}\\[1mm]
    $G(\mathrm{j}\omega_k)$
};


\node[
    box,
    draw=mimosblue,
    fill=mimosblue!8,
    text width=2.7cm
] (pre) at (3.0,0) {
    \textbf{FRF preprocessing}\\
    \textbf{and weighting}\\[1mm]
    {\footnotesize frequency range,}\\
    {\footnotesize channels, delay, weights}
};


\node[
    box,
    draw=mimosorange,
    fill=mimosorange!9,
    text width=2.7cm
] (init) at (6.6,0) {
    \textbf{Modal}\\
    \textbf{initialization}\\[1mm]
    $\omega_i^0,\ \zeta_i^0$
};


\draw[
    draw=mimosgreen,
    fill=mimosgreen!2,
    rounded corners=3pt,
    line width=0.8pt
]
    (8.80,-1.15)
    rectangle
    (16.75,1.35);

\node[
    text=mimosgreen!70!black,
    font=\footnotesize\bfseries,
    fill=white,
    inner xsep=3mm
] at (12.775,1.35) {
    Two-stage modal identification
};


\node[
    box,
    draw=mimosgreen,
    fill=mimosgreen!9,
    text width=2.55cm,
    minimum height=1.45cm
] (stage1) at (10.55,0) {
    \textbf{Stage 1}\\[0.7mm]
    {\footnotesize Additive estimation}\\[0.7mm]
    {\scriptsize LM / SRIVC}
};


\node[
    box,
    draw=mimosgreen,
    fill=mimosgreen!9,
    text width=2.55cm,
    minimum height=1.45cm
] (stage2) at (15.00,0) {
    \textbf{Stage 2}\\[0.7mm]
    {\footnotesize Modal estimation}\\[0.7mm]
    {\scriptsize LM}
};


\node[
    diamond,
    aspect=1.6,
    draw=mimosorange,
    fill=mimosorange!5,
    line width=0.7pt,
    minimum width=1.5cm,
    minimum height=1.05cm,
    inner sep=1pt,
    align=center,
    font=\footnotesize
] (accept) at (18.65,0) {
    Accept?
};


\node[
    align=center,
    text width=2cm
] (output) at (21.50,0) {
    \textbf{Modal model}\\[1mm]
    $P(s,\widehat{\rho})$
};


\node[
    box,
    draw=mimosorange,
    fill=mimosorange!5,
    text width=2.35cm,
    minimum height=0.85cm,
    font=\footnotesize
] (cmif) at (6.6,2) {
    \textbf{MIMOS workbench}\\
    inspect and refine
};


\draw[->, line width=0.7pt]
    (input.east) -- (pre.west);

\draw[->, line width=0.7pt]
    (pre.east) -- (init.west);

\draw[->, line width=0.7pt]
    (init.east) -- (stage1.west);

\draw[->, line width=0.7pt]
    (stage1.east) -- (stage2.west);

\node[
    font=\footnotesize,
    fill=mimosgreen!2,
    inner xsep=2mm,
    inner ysep=0.7mm
] at (12.775,0.38) {
    $\widehat{\beta},\,\hat\Sigma^{-1}_{\beta}$
};

\draw[->, line width=0.7pt]
    (stage2.east) --
    node[
        midway,
        above,
        font=\footnotesize
    ] {$\widehat{\rho}$}
    (accept.west);

\draw[->, line width=0.7pt]
    (accept.east) --
    node[
        midway,
        above,
        font=\footnotesize
    ] {Yes}
    (output.west);


\draw[
    <->,
    dashed,
    draw=mimosorange,
    line width=0.7pt
]
    (init.north) -- (cmif.south);


\draw[
    ->,
    dashed,
    draw=mimosorange,
    line width=0.7pt,
    rounded corners=3pt
]
    (accept.north)
    -- (18.65,2)
    -- node[
        midway,
        above,
        font=\footnotesize,
        text=mimosorange!80!black
    ] {No: reinitialize and restart}
    (cmif.east);

\end{tikzpicture}%
}

\caption{MIMOS identification workflow from estimated MIMO FRF 
to physical modal state-space model.}
\label{fig:mimos_workflow}
\end{figure*}

\subsection{Identification problem}\label{sec:id-problem}
The objective is to estimate the physical modal parameters
of \eqref{eq:modal-transfer} from estimated MIMO FRF data. Collect
the parameters of the rigid-body modes, flexible modes, and residual
term in the parameter vector
\begin{equation}\label{eq:modal-parameters}
\rho
=
\left[
\rho_{\mathrm{rb}}^\top,
\rho_{\mathrm{flex}}^\top,
\rho_{\mathrm{res}}^\top
\right]^\top,
\end{equation}
where $\rho_{\mathrm{flex}}$ contains the real parameters
of the proportional- or general-damping models in
\eqref{eq:proportional-mode} and \eqref{eq:general-mode},
respectively. Consider an estimated MIMO FRF
$G_k\in\mathbb C^{n_y\times n_u}$ at frequencies
$\omega_k$, $k=1,\ldots,N_\omega$. The modal identification problem is
formulated as the weighted frequency-domain least-squares problem
\begin{equation}\label{eq:modal-id-problem}
\widehat{\rho}
=
\underset{\rho}{\operatorname{arg\,min}}\;
\frac{1}{2N_\omega}
\sum_{k=1}^{N_\omega}
\left\|
\operatorname{vec}\!\left(
G_k-P(\mathrm{j}\omega_k,\rho)
\right)
\right\|_{W_k}^{2},
\end{equation}
where $W_k\in\mathbb C^{n_un_y\times n_un_y}$
is a frequency-dependent
weighting matrix. Here, $\operatorname{vec}(\cdot)$ stacks matrix columns into a vector
and $\|x\|_W^2=x^{\mathsf H}Wx$.
The problem \eqref{eq:modal-id-problem} is particularly
challenging due to the nonlinear dependence on modal frequencies
and damping ratios and the bilinear coupling of mode-shape
parameters across input--output channels, resulting in a
large-scale nonlinear estimation problem.

\section{Two-stage modal identification}\label{sec:two-stage}
To address the challenges of problem \eqref{eq:modal-id-problem},
MIMOS employs a two-stage identification approach, outlined in Figure \ref{fig:mimos_workflow}.
The first stage relaxes the bilinear mode-shape factorizations
by introducing unconstrained numerator matrices,
which are estimated directly from the FRF data.
The second stage subsequently projects the estimated additive model onto
the physical modal structure.

\subsection{Stage 1: Additive identification}\label{sec:stage1}
The first stage retains the second-order form of each modal contribution
while relaxing the bilinear mode-shape factorizations
by introducing unconstrained numerator matrices. This yields the additive model parameterization
\begin{equation}\label{eq:additive-model}
P_{\mathrm{a}}(s,\beta)
=
\frac{B_{\mathrm{rb}}}{s^2}
+
\sum_{i=1}^{n_{\mathrm{flex}}}
\frac{B_{i,1}s+B_{i,0}}
{s^2+a_{i,1}s+a_{i,0}}
+D_{\mathrm{res}},
\end{equation}
where
$B_{\mathrm{rb}},B_{i,1},B_{i,0}\in\mathbb R^{n_y\times n_u}$, with $B_{i,1}=0$ for proportional damping, and
$a_{i,1},a_{i,0}>0$ parameterize the denominator of the $i$th flexible
mode. The additive parameters are collected in the vector
\begin{equation}\label{eq:additive-parameters}
\beta
=
\left[
\beta_{\mathrm{rb}}^\top,
\beta_{\mathrm{flex}}^\top,
\beta_{\mathrm{res}}^\top
\right]^\top,
\end{equation}
where the components contain the rigid-body, flexible-mode, and
residual parameters of \eqref{eq:additive-model}, respectively.
The additive parameters are estimated by solving the weighted
frequency-domain least-squares problem
\begin{equation}\label{eq:stage1-problem}
\widehat{\beta}
=
\underset{\beta}{\operatorname{arg\,min}}\;
\frac{1}{2N_\omega}
\sum_{k=1}^{N_\omega}
\left\|
\operatorname{vec}\!\left(
G_k-P_{\mathrm{a}}(\mathrm{j}\omega_k,\beta)
\right)
\right\|_{W_k}^{2}.
\end{equation}

\subsection{Stage 2: Modal identification}\label{sec:stage2}
The second stage projects the additive estimate $\widehat{\beta}$ onto the
physical modal parameterization. The physical modal model defines a structured, lower-dimensional subset
of the additive parameter space through the mapping
\begin{equation}\label{eq:modal-map}
\beta=f(\rho),
\end{equation}
which maps the modal parameters in \eqref{eq:modal-parameters} to the
corresponding parameters of the additive representation in
\eqref{eq:additive-parameters}. The projection is weighted by an estimate of the information matrix corresponding to the
Stage~1 solution $\hat \beta$, which is computed as
\begin{equation}\label{eq:beta-information}
\hat \Sigma^{-1}_{\hat \beta}
=
J_{\hat \beta}^{\mathsf T}J_{\hat \beta},
\end{equation}
where $J_{\hat \beta}$ is the Jacobian of the weighted residual in the least-squares problem
\eqref{eq:stage1-problem} evaluated at $\widehat{\beta}$. The physical
modal parameters are then estimated by solving the least-squares problem
\begin{equation}\label{eq:stage2-problem}
\widehat{\rho}
=
\underset{\rho}{\operatorname{arg\,min}}\;
\frac{1}{2}
\left\|
\widehat{\beta}-f(\rho)
\right\|_{\hat \Sigma^{-1}_{\hat \beta}}^{2}.
\end{equation}
This information-weighted projection is an indirect prediction error
method \citep{Soderstrom1991AnIdentification}, in which the parameter-space criterion locally
approximates the frequency-domain identification criterion around
$\widehat{\beta}$. The resulting estimate $\widehat{\rho}$ defines the physical
modal model \eqref{eq:modal-transfer}, from which the
state-space realization \eqref{eq:modal-state-space}
is constructed directly.

\subsection{Advantages of the two-stage approach}
\label{sec:two-stage-properties}
The two-stage approach offers two main advantages over directly
solving \eqref{eq:modal-id-problem}. First, it reduces computational
complexity by separating the estimation of the input--output
dynamics from the enforcement of the physical modal structure.
Stage~1 estimates the additive model from the complete MIMO FRF
dataset, whereas Stage~2 performs the nonlinear mode-shape
factorization in the parameter space of $\widehat{\beta}$,
avoiding repeated evaluation of the frequency-domain criterion.
The information-weighted Stage~2 projection locally
approximates the original frequency-domain identification
criterion, allowing the asymptotic statistical properties
of direct estimation to be preserved.
The underlying indirect prediction error principle is
established in \citet{Soderstrom1991AnIdentification},
with a statistical analysis of the two-stage
approach provided in
\citet{Gonzalez2025StatisticallyIdentification}.

Second, the Stage~1 modal residue matrices enable
singular-value-based selection of the number $r_i$ of physical
modal components retained for each flexible mode.
The corresponding singular vectors provide essential
initial estimates of the mode-shape directions for
the nonconvex Stage~2 optimization problem.

\section{MIMOS identification workflow}\label{sec:toolbox}

MIMOS implements the two-stage identification method in the
workflow shown in Figure~\ref{fig:mimos_workflow}, integrating
FRF preprocessing and weighting, modal initialization, additive estimation,
and projection onto the physical modal structure.

\subsection{FRF preprocessing and weighting}\label{sec:preprocessing}
MIMOS allows the user to select the frequency ranges and
input--output channels used for identification.
A common time delay can be specified manually or estimated
automatically from the MIMO FRF, which is compensated in the data before estimation. The FRF residuals can be weighted using inverse-magnitude,
variance-based, or user-defined weighting.
Inverse-magnitude weighting emphasizes relative errors,
variance weighting accounts for the statistics of the
estimated FRFs, and user-defined weighting
allows emphasis on selected frequency ranges. The preprocessed FRFs and corresponding weighting filters
are used in the subsequent identification stages.

\subsection{Modal initialization}\label{sec:initialization}
\begin{figure}[b]
    \centering
    \includegraphics[width=\columnwidth]{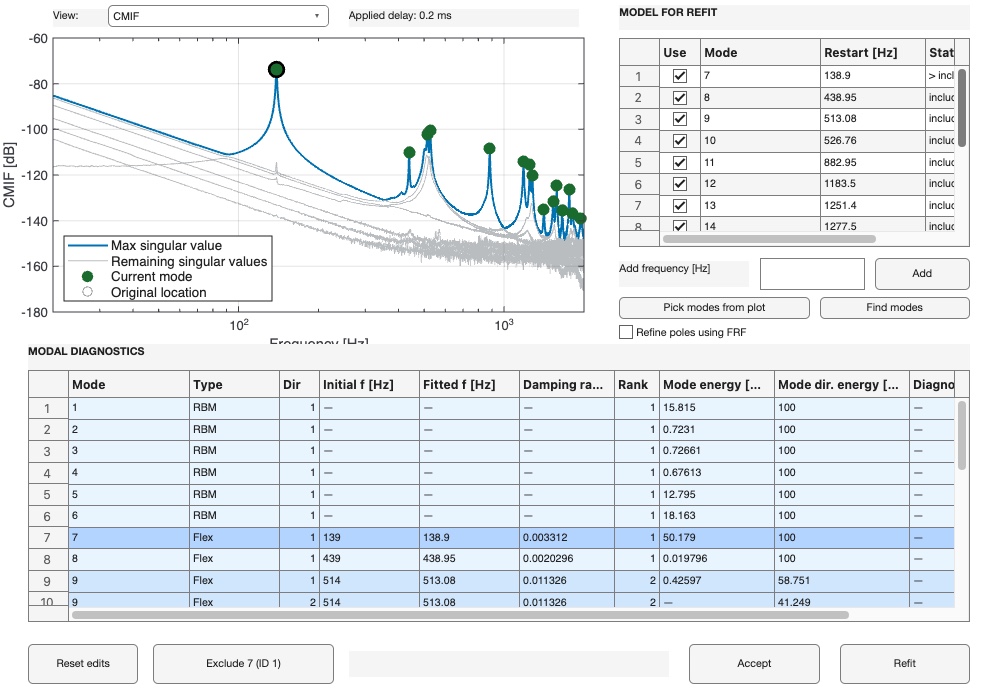}

\caption{MIMOS workbench GUI for interactive modal initialization,
inspection of modal diagnostics, and reinitialization and restarting
of the identification procedure.}

    \label{fig:initialization}
\end{figure}

Stage~1 requires initial estimates of the modal frequencies and
damping ratios. MIMOS automatically initializes these parameters
from the estimated MIMO FRF, using the complex mode indicator
function (CMIF) to detect candidate flexible modes
\citep{Shih1988ComplexEstimation}.
The initialization can be inspected and refined through the
interactive MIMOS workbench GUI shown in Figure~\ref{fig:initialization},
where modes can be added or removed and their frequencies
and damping ratios adjusted.
Alternatively, the initial modal parameters can be supplied manually.

\subsection{Stage 1: Additive estimation}\label{sec:toolbox-stage1}
The initialized natural frequencies and damping ratios define the
flexible-mode denominators in \eqref{eq:additive-model}.
With these denominators fixed, the unconstrained numerator matrices
are initialized using weighted linear least squares.
The complete additive parameter vector is subsequently refined
by solving \eqref{eq:stage1-problem} using Levenberg--Marquardt (LM),
with the additive SRIVC method introduced in
\citet{HulstModal2026} available as an alternative.
The estimate $\widehat{\beta}$ and information matrix
$\hat\Sigma_\beta^{-1}$ provide the input to Stage~2.

\begin{figure}[t]
    \centering
    \vspace{-2.4em}
    \includegraphics[width=\columnwidth]{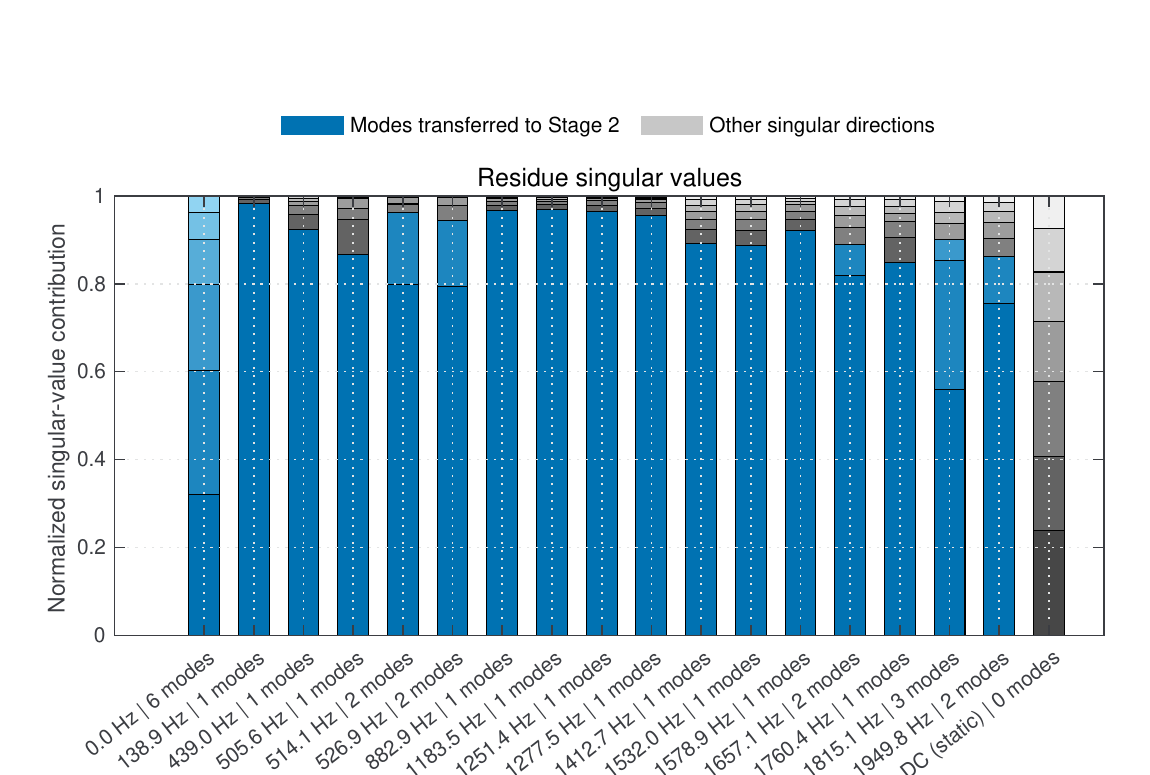}

\caption{Normalized singular-value contributions of the Stage~1
modal residues used to select the number $r_i$ of physical
components per flexible mode.}

    \label{fig:modal-svd}
\end{figure}

\subsection{Stage 2: Modal estimation}\label{sec:toolbox-stage2}
For each flexible mode, MIMOS selects the number $r_i$ of
physical modal components from the singular values of the
Stage~1 modal residues. For proportional damping, the residue matrix is $B_{i,0}$,
whereas for general damping the first-order pole residues
in \eqref{eq:general-mode} follow from partial-fraction
decomposition of the second-order term with numerator
$B_{i,1}s+B_{i,0}$ in \eqref{eq:additive-model}.
 The smallest $r_i$ capturing a user-specified percentage of the
squared singular-value content is retained.
The corresponding singular vectors initialize the input and
output mode-shape matrices.
Figure~\ref{fig:modal-svd} illustrates this selection.
The modal parameters are subsequently refined by solving
\eqref{eq:stage2-problem} using LM.
Alternatively, MIMOS supports direct FRF-based refinement,
with optional pole refinement.
The resulting estimate $\widehat{\rho}$ is used to construct
the modal state-space model in \eqref{eq:modal-state-space}.

\subsection{MIMOS workflow}\label{sec:mimos-usage}
MIMOS supports an iterative identification workflow in which
the modal model is fitted, inspected, and refined until a
satisfactory result is obtained.
The initial identification is performed using
\texttt{MIMOS.modalfit}, which automatically opens the
MIMOS workbench upon completion.
The GUI allows inspection of modal diagnostics, modification
of the mode selection and modal parameters, and repeated
refitting of both estimation stages.
The workbench can be reopened using \texttt{MIMOS.modalrefine}.
A minimal identification is configured
as follows.

\begin{lstlisting}[style=matlab]
% Identification settings
opts = MIMOS.defaultOptions();
opts.Model.DampingModel      = "proportional";
opts.Model.NumRigidBodyModes = 6;
opts.Data.FrequencyRange     = [20,2000];
opts.Data.Delay.TimeDelay    = "auto";

% Identification and interactive refinement
result = MIMOS.modalfit(G,opts);
\end{lstlisting}

Here, \texttt{G} contains the estimated MIMO FRF as a MATLAB
\texttt{frd} object, while \texttt{opts} specifies the
identification settings.
The returned \texttt{result} contains the identified physical
modal state-space model, modal parameters, fit quality,
and estimation diagnostics.

\section{Experimental case studies}\label{sec:experiments}

\begin{figure}[t!]
    \centering
    \begin{minipage}[c]{0.49\columnwidth}
        \centering
        \includegraphics[width=\linewidth]{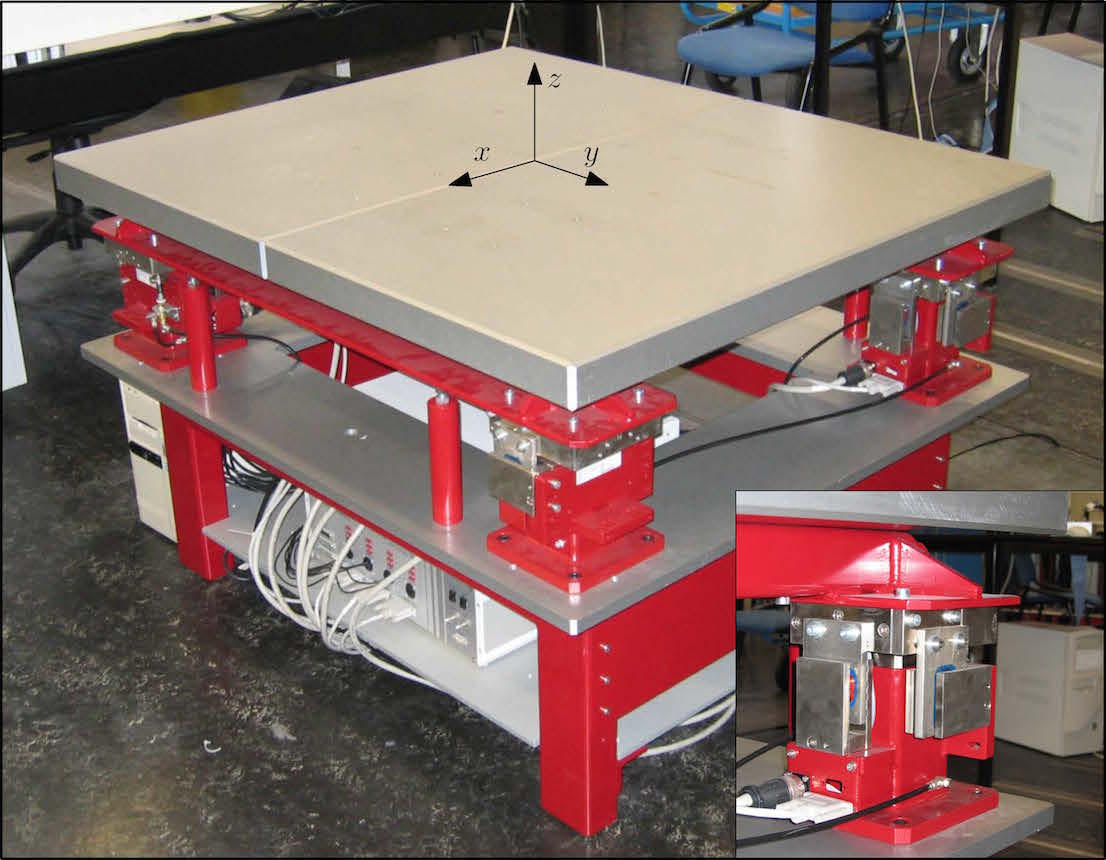}
    \end{minipage}\hfill
    \begin{minipage}[c]{0.49\columnwidth}
        \centering
        \includegraphics[width=\linewidth]{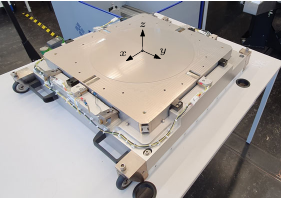}
    \end{minipage}
    \caption{Experimental systems used for validation:
    AVIS benchmark (left) and prototype wafer stage (right).}
    \label{fig:experimental-systems}
\end{figure}

MIMOS is evaluated on two experimental multivariable
mechatronic systems: the Active Vibration Isolation System
(AVIS) benchmark with open source data \citep{Voorhoeve2015Identification} and a prototype wafer stage, shown in
Figure~\ref{fig:experimental-systems}.
Table~\ref{tab:identification-results} summarizes the
results for proportional and general viscous damping.

\subsection{Active vibration isolation benchmark}\label{sec:avis}

The AVIS benchmark 
consists of eight voice-coil actuators and six geophones
measuring absolute velocity. The measured FRFs are
converted to displacement before identification.
MIMOS identifies 121 flexible modes for the $6\times8$
system using approximately $10^4$ frequency points.
For general viscous damping, Stage~1 estimates 11\,906
parameters and achieves a 94.10\% FRF fit.
Stage~2 reduces the model from 1452 to 834 states
and from 11\,906 to 7892 independent parameters,
while retaining a 93.45\% fit.
Figure~\ref{fig:avis_frf} compares the measured and
identified FRFs. In contrast, proportional damping
achieves only 67.35\% using the same initialization.
These results demonstrate the capability of MIMOS to identify large-scale MIMO modal models involving thousands of parameters and establish the importance of general viscous damping for accurately representing the AVIS dynamics.

\subsection{Prototype wafer stage}\label{sec:waferstage}
\begin{figure*}[t]
    \centering
    \includegraphics[width=0.95\textwidth]{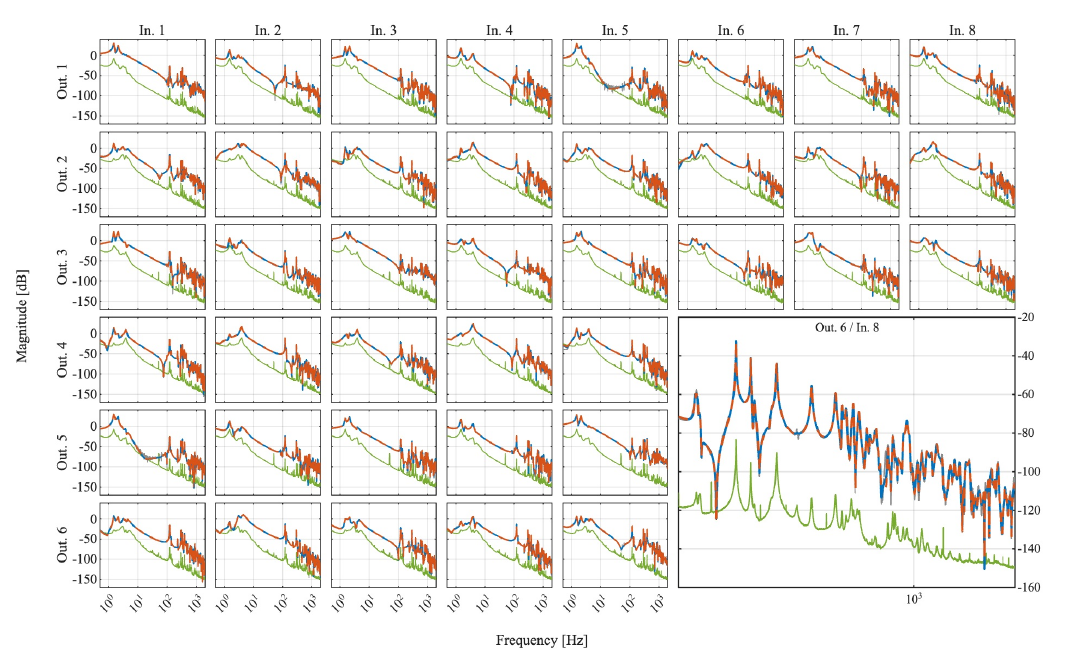}
    \par\vspace{-10pt}
    \caption{FRF magnitude plots for the $6\times8$ AVIS
    benchmark showing estimated FRFs \tikzline{mimosmeas}
    with standard deviation \tikzline{mimosnoise},
    Stage~1 additive model \tikzline{mimoss1},
    and Stage~2 modal model \tikzline{mimoss2},
    with an enlarged view of channel Out.~6 / In.~8.}
    \label{fig:avis_frf}

    \includegraphics[width=0.95\textwidth]{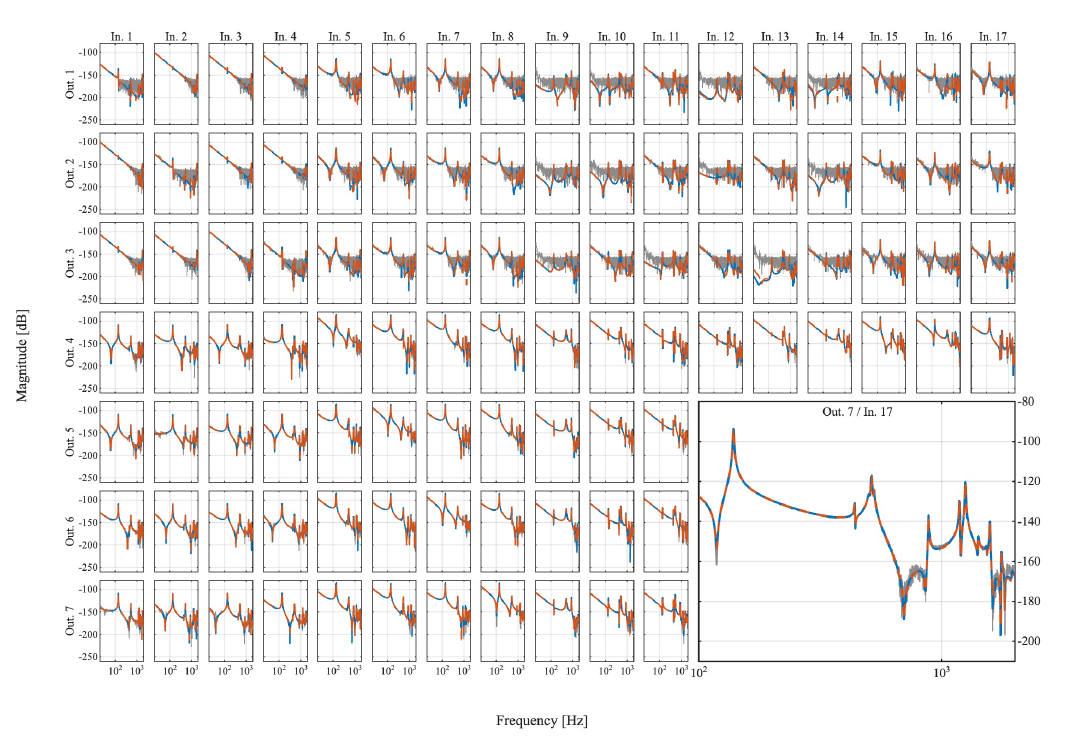}
    \par\vspace{-10pt}
    \caption{FRF magnitude plots for the $7\times17$
    prototype wafer stage showing estimated FRFs
    \tikzline{mimosmeas}, Stage~1 additive model
    \tikzline{mimoss1}, and Stage~2 modal model
    \tikzline{mimoss2}, with an enlarged view of
    channel Out.~7 / In.~17.}
    \label{fig:waferstage_frf}
\end{figure*}

The prototype wafer stage
\citep{Voorhoeve2021IdentifyingStage,HulstModal2026}
has 17 actuator inputs and seven position measurements.
MIMOS identifies six rigid-body and 20 flexible modes
for the $7\times17$ system using approximately
$4\times10^3$ frequency points.
For proportional and general viscous damping,
Stage~2 reduces the state-space order from 288 to
84 and 90 and the independent parameter count
by 61.1\% and 62.9\%, respectively.
The corresponding FRF fits decrease by only
0.09 and 0.16 percentage points relative to Stage~1,
reaching 81.66\% and 82.85\%, with Stage~2
computation times of 3.4~s and 4.3~s.
Figure~\ref{fig:waferstage_frf} compares the measured
and identified FRFs. These results demonstrate
the capability of MIMOS to achieve substantial
structural reduction while closely preserving
the identified input--output dynamics.


\begin{table}[!b]
\centering
\caption{AVIS and OAT identification results.}
\label{tab:identification-results}
\setlength{\tabcolsep}{1.3pt}
\scriptsize
\begin{tabular}{@{}llcccccrrrr@{}}
\toprule
System & Damp. &
$n_y\times n_u$ &
$N_\omega$ &
$n_{\mathrm{rb}}$ &
$n_{\mathrm{flex}}$ &
Stage &
$n_x$ &
$n_p$ &
Fit [\%] &
$t$ [s] \\
\midrule
AVIS & Prop. & $6\times8$ & $10^4$ & 0 & 121
& 1 & 1452 & 6098 & 68.45 & 261.1 \\
& & & & & &
2 & 800 & 4302 & 67.35 & 29.5 \\
& Gen. & $6\times8$ & $10^4$ & 0 & 121
& 1 & 1452 & 11906 & 94.10 & 381.5 \\
& & & & & &
2 & 834 & 7892 & 93.45 & 67.6 \\
\midrule
Stage & Prop. & $7\times17$ & $4\cdot10^3$ & 6 & 20
& 1 & 288 & 2658 & 81.75 & 8.1 \\
& & & & & &
2 & 84 & 1033 & 81.66 & 3.4 \\
& Gen. & $7\times17$ & $4\cdot10^3$ & 6 & 20
& 1 & 288 & 5038 & 83.01 & 8.7 \\
& & & & & &
2 & 90 & 1869 & 82.85 & 4.3 \\
\bottomrule
\end{tabular}
\end{table}

\section{Conclusions}\label{sec:conclusion}

This paper presented MIMOS, an open-source MATLAB toolbox for
identifying physically structured modal models from multivariable
FRF data. MIMOS enables practical identification of accurate, minimal-order,
and physically interpretable models from large FRF datasets,
providing a basis for analyzing system dynamics and model-based
control design. Experimental results on two industrial mechatronic systems
demonstrate the capability to handle thousands of parameters,
accommodate general viscous damping, and substantially reduce model
complexity while closely preserving the identified input--output
dynamics.

\bibliography{references}
\end{document}